\documentclass{emulateapj}
\usepackage{epsf}
\usepackage{apjfonts}

\slugcomment{To appear in ApJ (submitted March 9, 2004)}

\newcommand\Rein{R_{\rm ein}}
\newcommand\env{{\rm env}}
\newcommand\xx{{\bf x}}
\newcommand\uu{{\bf u}}
\newcommand\defl[1]{\left|\nabla\phi(\xx_{#1})\right|}
\newcommand\Bsig{{\tilde\sigma}}

\newcommand\refeq[1]{eq.~(\ref{eq:#1})}
\newcommand\refsec[1]{\S~\ref{sec:#1}}
\newcommand\reffig[1]{Figure~\ref{fig:#1}}
\newcommand\reffigs[2]{Figures~\ref{fig:#1} and \ref{fig:#2}}
\newcommand\reftab[1]{Table~\ref{tab:#1}}

\begin{document}

\title{The Importance of Lens Galaxy Environments}

\author{Charles R.\ Keeton\altaffilmark{1}} 
\affil{Astronomy \& Astrophysics Department, University of Chicago,
5640 S.\ Ellis Ave., Chicago, IL 60637}

\author{Ann I.\ Zabludoff}
\affil{Steward Observatory, University of Arizona, 933 N.\ Cherry Ave.,
Tucson, AZ 85721}

\altaffiltext{1}{Hubble Fellow}

\begin{abstract}

It is suspected that many strong gravitational lens galaxies lie
in poor groups or rich clusters of galaxies, which modify the
lens potentials.  Unfortunately, little is actually known about
the environments of most lenses, so environmental effects in
lens models are often unconstrained and sometimes ignored.  We
show that such poor knowledge of environments introduces
significant biases and uncertainties into a variety of lensing
applications.  Specifically, we create a mock poor group of 13
galaxies that resembles real groups, generate a sample of mock
lenses associated with each member galaxy, and then analyze the
lenses with standard techniques.  We find that standard models
of 2-image (double) lenses, which neglect environment, grossly
overestimate both the ellipticity of the lens galaxy
($\Delta e/e \sim 0.5$) and the Hubble constant
($\Delta h/h \sim 0.22$).  Standard models of 4-image (quad)
lenses, which approximate the environment as a tidal shear,
recover the ellipticity reasonably well
($|\Delta e/e| \lesssim 0.24$) but overestimate the Hubble
constant ($\Delta h/h \sim 0.15$), and have significant
($\sim$30\%) errors in the millilensing analyses used to constrain
the amount of substructure in dark matter halos.  For both doubles
and quads, standard models slightly overestimate the velocity
dispersion of the lens galaxy ($\Delta\sigma/\sigma \sim 0.06$),
and underestimate the magnifications of the images
($\Delta\mu/\mu \sim -0.25$).  Standard analyses that use the
statistics of lens populations to place limits on the dark
energy overestimate $\Omega_\Lambda$ (by 0.05--0.14), and
underestimate the ratio of quads to doubles (by a factor of 2).
The systematic biases related to environment help explain some
long-standing puzzles (such as the high observed quad/double
ratio), but aggravate others (such as the low value of $H_0$
inferred from lensing).  Most of the biases are caused by neglect
of the convergence (gravitational focusing) from the mass
associated with the environment, but additional uncertainty is
introduced by neglect of higher-order terms in the lens potential.
Fortunately, we show that directly observing and modeling lens
environments should make it possible to remove the biases and
reduce the uncertainties.  Such sophisticated lensing analyses
will require finding the other galaxies that are members of the
lensing groups, and measuring the group centroids and velocity
dispersions, but they should reduce systematic effects associated
with environments to the few percent level.

\end{abstract}


\section{Introduction}

The study of strong gravitational lenses offers unique constraints
on the masses and properties of galaxy dark matter halos
\citep*[e.g.,][]{csk91,kkf,rkk,tk04},
the properties of quasars \citep[e.g.,][]{nemiroff,richards}
and their host galaxies \citep*[e.g.,][]{rix01,kkm,peng},
the Hubble constant \citep[e.g.,][]{refsdal,ks},
the nature of dark matter \citep[e.g.,][]{metcalf,dk},
and the properties of dark energy
\citep[e.g.,][]{turner90,csk96,chae,mitchell,linder}.
The number of strong lenses is approaching 100, with useful
subsamples that number in the tens \citep*[e.g.,][]{CLASS,rusin,ofek},
and will grow dramatically with ongoing and future surveys
\citep*[e.g.,][]{kuhlen}.  The relative positions and fluxes of
the lensed images are routinely measured to high precision with
the Hubble Space Telescope and radio interferometers
\citep*[e.g.,][]{lehar,patnaik,trotter}; Einstein rings or arcs
are proving to be common in high-resolution near-IR images
\citep[e.g.,][]{kkm}; and image time delays and lens galaxy
velocity dispersions are succumbing to concerted observational
effort
\citep[e.g.,][]{1115tdel,burud02a,burud02b,1608tdel,colley,tk2016,tk04}.
With such extensive and high-quality data in hand, the results of
lensing analyses are limited mainly by systematic uncertainties in
the lens models required to interpret the data.

We are interested in systematic uncertainties related to the fact
that lens galaxies are not isolated.  Many lenses are produced by
early-type galaxies that reside in overdense regions like poor
groups or rich clusters of galaxies \citep[e.g.,][]{young,kundic,
kundic1422,tonry,tonry0751,kneib,fassnacht,iva,kurtis}.  The
additional mass near the lens contributes both a ``convergence''
(additional gravitational focusing) and ``shear'' (gravitational
tidal force)\footnote{This shear is formally equivalent to that
probed with weak lensing analyses of clusters and cosmic shear
\citep[see reviews by][]{WL,cos-shear}, even though it is detected
in a different way.} to the lens potential, plus higher-order terms
that may or may not be small (see \refsec{mockgrp}).  One set of
systematic uncertainties in lens models occurs because the
convergence is degenerate with the lens galaxy mass, in all lens
observables but the time delays; this is the well-known ``mass-sheet
degeneracy'' \citep*{gorenstein,saha}.  Because of the degeneracy,
the convergence is usually omitted from lens models, which leads
to a variety of biases.  Attempts to correct the biases --- for
example, by using independent weak lensing analyses to constrain
the convergence \citep[e.g.,][]{bf0957} --- have been rare.

A second set of systematic effects occurs because the shear is
approximately degenerate with the ellipticity of the lens galaxy
\citep*{kks}.  Two-image (double) lenses suffer badly: even when
environmental effects are strong, models without shear fit the
data just fine, and models that include shear reveal it to be
degenerate with ellipticity.  Without independent knowledge of
the environment, the natural choice for doubles is to omit shear,
but that must introduce errors into the models.  The situation is
better for four-image (quad) lenses, because the degeneracy between
ellipticity and shear is only approximate.  Both effects produce
quadrupole terms in the lens potential, but with different radial
dependences: $\phi \propto \gamma\,r^2 \cos2(\theta-\theta_\gamma)$
for shear, and $\phi \propto e\,r \cos2(\theta-\theta_e)$ for
ellipticity.  Quads generally have enough constraints to detect
the difference, and thus to reveal that shear cannot be neglected
\citep[e.g.,][]{kks}.  Even so, models of quads may not uniquely
constrain both the ellipticity and shear, leaving a large range of
allowed models.  Furthermore, for both doubles and quads we must
ask whether approximating the environmental effects as a simple
shear (neglecting higher-order terms in the lens potential) is
adequate.

These issues can be collected into a general question:
{\it If lens galaxy environments are poorly known, how wrong will
standard lensing analyses be?\/}  We address the question by placing
galaxies in simple but realistic environments whose properties we
understand and control, and using them to generate catalogs of mock
lenses.  We then apply standard lensing analyses to the mock lenses,
and study whether the results accurately recover the input parameters.
We consider a wide range of astrophysical and cosmological problems
to which lensing is applied, including the shapes and masses of
galaxy dark matter halos, the amount of substructure in dark matter
halos, the properties of lensed sources, the Hubble constant, and
the dark energy density.  In this paper we begin by examining a
single (but typical) case to identify problems that arise when
lens environments are unknown.  In a subsequent analysis we will
study in more detail how the errors depend on the properties of
the environment.

We stress that we focus on arcsecond-scale lens systems in which
the lens potential is dominated by one galaxy \citep[or occasionally
two or three galaxies in close proximity;][]{1359,0134}, and the
contribution from the environment can be treated as a perturbation.
This is the case for nearly all multiply-imaged quasars and radio
sources.  It differs from the case of lensed arcs
\citep[e.g.,][]{gladders,zaritsky03} and wide-separation lensed
quasars \citep[e.g.,][]{1004}, which are produced by the large dark
matter halos of clusters.

In this paper we focus on systematic effects associated with the
angular structure of the potential.  There is an additional set of
systematic effects associated with the fact that in models of many
lenses the radial density profile of the lens galaxy is degenerate.
We explicitly neglect this degeneracy by always assuming that the
galaxies can be modeled as isothermal ellipsoids.  This simple
assumption seems to be remarkably good \citep[e.g.,][]{zaritsky,
rix97,gerhard,mckay,tk2016,kt1608,rkk,sheldon}.  Even if this
assumption comes into question \citep[see][]{romanowsky}, the
crucial point is that we use a fixed radial density profile for
both generating and modeling mock lenses.  In this way we avoid
any systematic errors associated with the radial profile, and
highlight systematic effects associated with environment.  In the
follow-up analysis we will also examine systematics related to
the radial profile, and consider whether knowledge of lens galaxy
environments can actually help break the profile degeneracy.  Other
recent investigations have studied the radial profile degeneracy
in considerable detail \citep[e.g.,][]{tk2016,tk1115,tk04,rkk}.

This paper is organized as follows.  We first create a typical
lens environment, a poor group of galaxies that mimics the group
around the observed lens PG~1115+080 (\refsec{mockgrp}).  We use
the galaxies in the group to create a catalog of mock lenses
(\refsec{mocklens}).  We then analyze the mock lenses with
standard techniques to identify environment-induced uncertainties
and biases in a variety of lensing applications (\refsec{problems}).
Finally, we show how knowledge of the environment can be used to
remove the systematic effects (\refsec{fix}).  We offer some
general comments in \refsec{disc}, and summarize our conclusions
in \refsec{concl}.

\section{A mock group of galaxies}
\label{sec:mockgrp}

\begin{figure*}
\plottwo{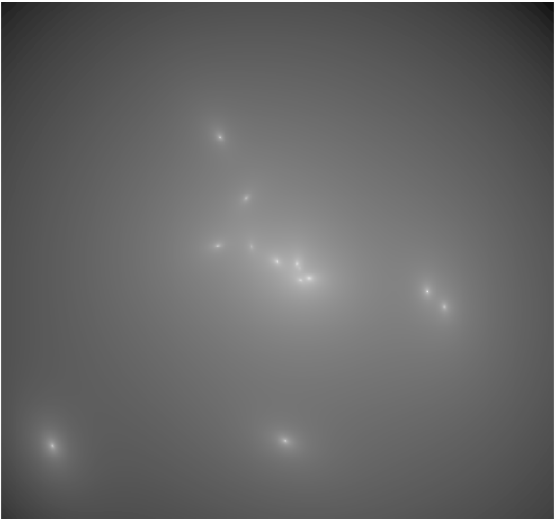}{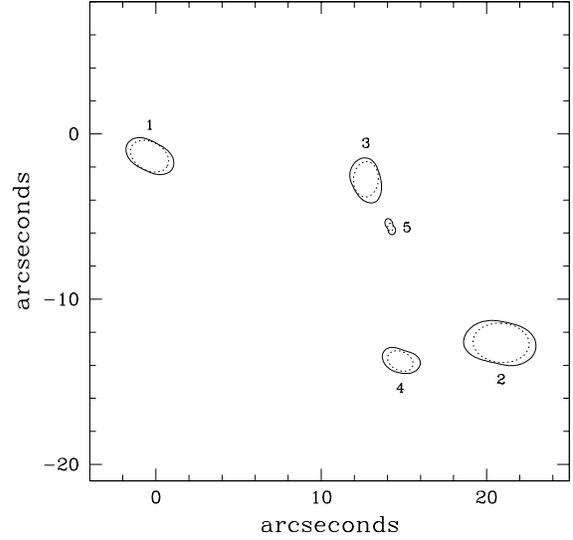}
\caption{
(Left) Logarithmic surface mass density map for the mock group, in
the model with all the mass associated with the galaxies and their
dark matter halos.  The map is 6\arcmin\ (570 $h^{-1}$ kpc) on a side.
(Right) Lensing critical curves in the central region of the group.
The solid curves show the actual critical curves; for comparison,
the dotted curves show the critical curve each galaxy would have if
it were isolated.  The galaxies are labeled with the indices from
\reftab{grp}.
}\label{fig:kmap}
\end{figure*}

Although the full distribution of lens galaxy environments is not
well known, predictions \citep*{kcz} and observations
\citep[e.g.,][]{kundic,kundic1422,tonry,tonry0751,fassnacht,iva,kurtis}
indicate that the most common lens environments are probably poor
groups of galaxies.  We seek to create a mock group that mimics
the group at redshift $z=0.31$ around the 4-image lens PG~1115+080
\citep{kundic,tonry,iva}.  This group appears to be typical of lens
environments, and to be similar to X-ray selected groups in the
nearby universe \citep{ZM98,ZM00} in terms of its galaxy population,
kinematics, and X-ray properties \citep{grant,iva}.  In addition,
this lens is a favorite for many lensing applications
\citep[e.g.,][]{1115tdel,kk1115,saha97,impey,zhao1115,tk1115},
so it offers an insightful example.

The lens in PG~1115+080 consists of four images of a quasar at
redshift $z_s=1.72$ around an early-type galaxy at redshift
$z_l=0.31$.  \citet{kundic} and \citet{tonry} discovered that
there are four other galaxies at the same redshift spanning
24\arcsec\ on the sky and forming a group with velocity dispersion
$\sigma \sim 300$~km~s$^{-1}$.  \citet{iva} have expanded the
group membership to 13 galaxies spanning 4.5\arcmin, and calculated
a velocity dispersion of $\sigma = 354\pm53$~km~s$^{-1}$.  We use
the relative positions of these galaxies, together with the
relative magnitudes given by \citet{kurtis}.

In deciding how to distribute the mass within the group, it is
useful to consider two extreme cases that bound the possibilities.
One case places all of the mass in the group member galaxies and
their individual dark matter halos; this could represent a system
where the galaxies have not interacted tidally with each other.
The other case places all of the mass in a common dark matter halo,
such that the galaxies are essentially massless tracers of the
group potential.  In nearby groups, the kinematics of group member
galaxies and the presence of extended X-ray luminous halos suggest
that the most of the mass is in a common group halo
\citep{ZM98,MZ98}.  The situation is less clear for the distant
groups ($0.3 \lesssim z \lesssim 1$) where lens galaxies are likely
to be found.  X-ray halos have been detected in two lensing groups
at $z \sim 0.3$ \citep[including PG~1115+080;][]{grant}, but
studying them well enough to probe the group mass distributions is
exceedingly difficult (the X-ray halos are faint extended sources
that have the bright quasar image superposed).  Theoretically, one
might expect distant groups to be less dynamically evolved than
nearby groups, and thus to have less mass in a common halo.  Given
the uncertainties, we have analyzed both of these extreme cases.
The systematic errors in lens models are qualitatively quite similar
for the two cases, so we present only the case where all of the
mass is in the galaxies and their individual dark matter halos.
This choice makes it easier to discuss how observations of lens
environments can be used to reduce or remove the model errors (see
\refsec{fix}).

We model the galaxies as isothermal ellipsoids, which is
consistent with evidence from strong lensing, stellar dynamics,
and X-ray studies on small scales
\citep{fabbiano,rix97,gerhard,tk2016,kt1608,rkk},
and with galaxy-galaxy lensing and satellite kinematics on large
scales \citep{zaritsky,mckay,sheldon}.  An isothermal ellipsoid
has projected surface mass density (in units of the critical
density for lensing)
\begin{equation} \label{eq:sie}
  \kappa = \frac{\Sigma}{\Sigma_{\rm crit}}
  = \frac{b}{2r} \left[\frac{1+q^2}
    {(1+q^2)+(1-q^2)\cos2(\theta-\theta_0)}\right]^{1/2} ,
\end{equation}
where $q \le 1$ is the axis ratio, $\theta_0$ is the orientation
angle (defined as a position angle measured East of North), and
$b$ is a mass parameter related to the velocity dispersion $\sigma$
by
\begin{equation} \label{eq:b}
  b = 4\pi \left(\frac{\sigma}{c}\right)^2 \frac{D_{ls}}{D_{os}}\ ,
\end{equation}
where $D_{os}$ and $D_{ls}$ are angular diameter distances from
the observer to the source and from the lens to the source,
respectively.  For a spherical galaxy $b$ equals the Einstein
radius, while for a nonspherical galaxy $b$ and $\Rein$ differ
by no more than a few percent for reasonable axis ratios.  The
lensing properties of isothermal ellipsoids are given by
\citet{kassiola}, \citet*{kormann}, and \citet{spirals}.

\begin{deluxetable*}{crrccrcccc}
\tablecaption{Properties of the Mock Group\label{tab:grp}}
\tablehead{
  \colhead{Galaxy} &
  \colhead{$x$ (\arcsec)} &
  \colhead{$y$ (\arcsec)} &
  \colhead{$b$ (\arcsec)} &
  \colhead{$e$} &
  \colhead{$\theta_0$ (\arcdeg)} &
  \colhead{$\kappa_{\env}$} &
  \colhead{$\gamma_{\env}$} &
  \colhead{$N_2$} &
  \colhead{$N_4$}
}
\startdata
 1 & $   0.0$ & $   0.0$ & 1.00 & 0.3 & $ 60.3$ & 0.1123 & 0.0950 & 422 & 215 \\
 2 & $  21.3$ & $ -11.3$ & 1.38 & 0.3 & $ 85.6$ & 0.1440 & 0.0859 & 931 & 415 \\
 3 & $  13.1$ & $  -1.4$ & 0.87 & 0.3 & $ -3.3$ & 0.1782 & 0.1090 & 419 & 145 \\
 4 & $  15.2$ & $ -12.4$ & 0.66 & 0.3 & $ 63.9$ & 0.2438 & 0.0797 & 293 & 138 \\
 5 & $  14.6$ & $  -4.3$ & 0.19 & 0.3 & $  1.5$ & 0.3148 & 0.2428 &  11 &  37 \\
 6 & $   5.5$ & $-117.4$ & 0.43 & 0.3 & $ 66.4$ & 0.0268 & 0.0209 &  70 &  11 \\
 7 & $ -38.6$ & $  10.4$ & 0.39 & 0.3 & $-72.1$ & 0.0626 & 0.0449 &  72 &  17 \\
 8 & $ -20.2$ & $  41.1$ & 0.28 & 0.3 & $-34.7$ & 0.0589 & 0.0439 &  35 &   6 \\
 9 & $  97.9$ & $ -19.5$ & 0.51 & 0.3 & $  9.0$ & 0.0476 & 0.0382 & 123 &  16 \\
10 & $ -37.3$ & $  81.4$ & 0.27 & 0.3 & $ 28.1$ & 0.0337 & 0.0300 &  37 &   6 \\
11 & $ 109.4$ & $ -29.8$ & 0.41 & 0.3 & $ 11.2$ & 0.0469 & 0.0388 &  74 &  16 \\
12 & $ -16.5$ & $   9.6$ & 0.35 & 0.3 & $ 11.6$ & 0.0983 & 0.0814 &  62 &  12 \\
13 & $-146.4$ & $-120.5$ & 0.35 & 0.3 & $ 22.7$ & 0.0168 & 0.0149 &  57 &   9 \\
\enddata
\tablecomments{
Each galaxy has position $(x,y)$, ellipticity and position angle
$(e,\theta_0)$, and mass parameter $b$; and feels a convergence
$\kappa_{\env}$ and shear $\gamma_{\env}$ from its environment;
$N_2$ and $N_4$ give the number of mock double and quad lenses in
our catalog that are associated with each galaxy.
}
\end{deluxetable*}

The spectra of group member galaxies indicate that many are
early-type galaxies \citep{iva}, so we use the Faber-Jackson
relation ($L \propto \sigma^4$) to estimate the galaxies' velocity
dispersions and $b$ parameters.  Specifically, the ratio of the
$b$ parameters for two galaxies $i$ and $j$ is estimated from
their relative magnitudes,
\begin{equation} \label{eq:brat}
  \frac{b_i}{b_j} = 10^{-0.2(m_i-m_j)} .
\end{equation}
Scatter in the Faber-Jackson relation \citep[e.g.,][]{sheth} could
cause scatter between these $b$ ratios and the actual values for
the PG~1115+080 group, but that is irrelevant because we use
\refeq{brat} to {\it define\/} the mock group.  We determine the
scale by setting $b=1\farcs0$ for the galaxy that is an analog of
the lens galaxy in PG~1115+080 (galaxy \#1 in \reftab{grp} below);
this value is both consistent with the observed lens \citep{impey}
and typical of all lenses
\citep[see the image separation distributions in, e.g.,][]{kcz,kuhlen}.

We assign all of the galaxies an ellipticity $e=1-q=0.3$, which
is typical for early-type galaxies \citep[e.g.,][]{bender,saglia,
jorgensen}, and we give them random orientations.  The resulting
parameters for the mock group galaxies are given in \reftab{grp}.
A visual sense of the group is given in \reffig{kmap}.  Note that
the left panel shows the projected mass (not light) distribution,
so it is perhaps more comparable to maps from $N$-body simulations
than to images of observed groups.

It is worth pointing out that even though we study a single global
system (the group), each galaxy has a unique local environment so
we actually have 13 sample lens galaxy environments.  As a further
test of whether the group is reasonable, we can quantify the
range of environmental contributions to the lens potential.  For
galaxy $i$, the environmental piece of the potential is the part
due to all the other galaxies:
\begin{equation}
  \phi = \phi_{i} + \phi_{\env,i}\,,
  \quad
  \phi_{\env,i} \equiv \sum_{j \ne i} \phi_{j}\,.
\end{equation}
The lensed images tend to lie near a radius (the Einstein radius)
that is small compared with the typical distance between galaxies.
Therefore we might expand $\phi_{\env,i}$ as a Taylor series in
polar coordinates centered on galaxy $i$.  The lowest order
significant terms are\footnote{The ${\cal O}(r^0)$ term
represents the unobservable zero point of the potential, while
the ${\cal O}(r)$ terms represent an unobservable uniform
deflection.}
\begin{equation} \label{eq:env}
  \phi_{\env} = \frac{r^2}{2}\Bigl[ \kappa_{\env}
    + \gamma_{\env}\cos2(\theta-\theta_{\gamma}) \Bigr]
    + {\cal O}(r^3)\,,
\end{equation}
where $\kappa_{\env}$ and $\gamma_{\env}$ represent, respectively,
the convergence (gravitational focusing) and shear (tidal
perturbation) discussed in the Introduction.  Equivalently, the
convergence and shear can be expressed in terms of derivatives of
the lens potential \citep*[e.g.,][]{SEF}:
\begin{eqnarray}
  \kappa_{\env} &=& \frac{1}{2}\left(
    \frac{\partial^2\phi_{\env}}{\partial x^2} +
    \frac{\partial^2\phi_{\env}}{\partial y^2} \right)\,, \label{eq:k} \\
  \gamma_{\env} &=& \left( \gamma_{\env,+}^2 +
    \gamma_{\env,\times}^2 \right)^{1/2}\,, \label{eq:g} \\
  \gamma_{\env,+} &=& \frac{1}{2}\left(
    \frac{\partial^2\phi_{\env}}{\partial x^2} -
    \frac{\partial^2\phi_{\env}}{\partial y^2} \right)\,, \\
  \gamma_{\env,\times} &=&
    \frac{\partial^2\phi_{\env}}{\partial x \partial y}\ .
\end{eqnarray}
Even if the higher-order terms in \refeq{env} are not negligible,
the convergence and shear are still a useful way to quantify the
environmental contribution to the lens potential.

\begin{figure}
\plotone{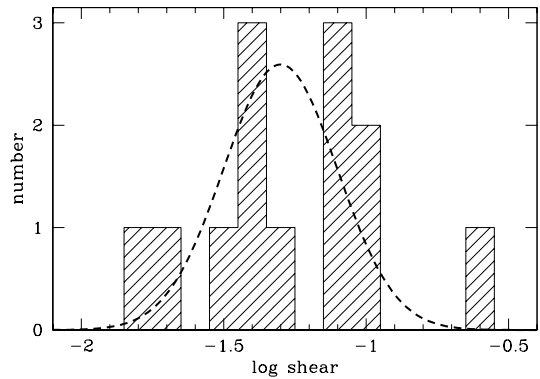}
\caption{
The histogram shows the shear distribution for the 13 galaxies in
the mock group.  The dashed curve shows the distribution for
early-type galaxies in galaxy formation simulations \citep{holder}.
}\label{fig:gdist}
\end{figure}

\reftab{grp} lists the convergence and shear computed for each
galaxy in the mock group, and \reffig{gdist} shows a histogram of
the shears.  For comparison, \citet{holder} have computed the
distribution of shears at the positions of early-type galaxies
in $N$-body and semi-analytic models of galaxy formation.  This
distribution, which is also shown in \reffig{gdist}, represents
the range of shears expected in standard galaxy formation models,
and is also broadly consistent with the shears required to fit
observed 4-image lenses \citep[see][]{kks,holder}; so it provides
at least a rough guide to the shears felt by real lenses.
Although the sample size is small, the shear distribution for our
mock group is remarkably similar to the distribution found by
Holder \& Schechter.\footnote{The point in the high-shear tail
is galaxy \#5, a small galaxy (near the center of \reffig{kmap})
that lies very close to, and is strongly perturbed by, galaxy \#3.
This galaxy has a small lensing cross section, so even if its
shear is unusually large it contributes little to our analysis
and does not affect our conclusions.}  While this test is by no
means conclusive, it does suggest that our mock group is not way
off base and is in fact quite compatible with the expected
distribution of environmental contributions to lens potentials.

\section{A mock lens catalog}
\label{sec:mocklens}

\subsection{Basic approach}

While our mock group represents a single global system, the 13
member galaxies all have different {\it local\/} environments and
therefore sample the range of environments that may be common to
group galaxies.  We have already made use of this fact in
examining the distribution of shears and using it to argue that
our group is reasonable.  The next step is to sample the range of
lens systems associated with these different environments.  We do
this by creating a catalog of mock lenses: we place many random
sources behind the group, determine which are strongly lensed,
and tabulate the positions, magnifications, and time delays of
all the lensed images.

We then employ the mock lens catalog to test the effects of
environment in two ways.  First, we imagine that we have observed
the mock lenses, and we fit each one individually with standard
mass models to constrain the lens galaxy properties (mass,
ellipticity, substructure, etc.) and Hubble constant $H_0$.
Second, we use the statistics of the lens sample to constrain
the cosmological constant $\Omega_\Lambda$ and to study the
relative numbers of quad and double lenses.  The main question
is whether or not standard lensing analyses, in which environment
is unknown or poorly understood, yield accurate constraints on
the various parameters.  In essence, we are using a Monte Carlo
approach to sample the errors in lensing analyses that are
associated with group environments.  The remainder of this
section is devoted to the details of the Monte Carlo techniques.

\subsection{Application to individual lens systems}
\label{sec:mock1}

In creating the mock lens catalog, solving the lens equation to
find the images of a given source is straightforward using the
algorithm and software by \citet{lenscode}.  The only subtleties
relate to how the random background sources are selected.  To
sample lens properties realistically, we must account for
``magnification bias,'' or the fact that highly magnified lens
systems are more readily discovered than less magnified systems
\citep*[e.g.,][]{TOG}.  If we want to mimic real samples, and in
particular if we want to avoid undersampling quad lenses, when
choosing sources we should give more weight to those with large
magnifications.  This amounts to applying a non-uniform probability
density in the source plane, which has the form
\begin{equation} \label{eq:psrc1}
  p(\uu)\,d\uu = f(\uu)\
    \frac{N(S_0/\mu(\uu))}{N(S_0)}\ d\uu\,,
\end{equation}
where $f(\uu)$ is 1 if a source at position $\uu$ is
multiply-imaged and 0 otherwise, $\mu(\uu)$ is the magnification
of a source at that position, $N(S)$ is the number of background
sources brighter than flux $S$, and $S_0$ is the flux limit of a
survey.  If the luminosity function of sources has a power law
form, $dN/dS \propto S^{-\nu}$, then the probability density
simplifies to
\begin{equation} \label{eq:psrc2}
  p(\uu)\,d\uu = f(\uu)\,\mu(\uu)^{\nu-1}\ d\uu\,.
\end{equation}
Formally, we just draw from this probability density to generate
the sample of sources, and then solve the lens equation to obtain
the mock lens catalog.

In practice, the probability density \refeq{psrc2} is quite
complicated (because of the lensing caustics, for example), and
there is no general algorithm for drawing from a complicated
two-dimensional probability density.  Fortunately, a simplification
is possible.  In the special case $\nu=2$ we can transform the
probability density as follows:
\begin{equation} \label{eq:pimg}
  p(\uu)\,d\uu
  = f(\uu) \sum_{i=1}^{N_{\rm img}}
    \left|\frac{\partial\xx_i}{\partial\uu}\right| d\uu
  = f(\xx)\,d\xx\,,
\end{equation}
where $f(\xx)$ is 1 if an image at position $\xx$ corresponds to
a source that is multiply-imaged, and 0 otherwise.  The first
equality uses the fact that the total magnification is the sum of
the magnifications for all the images, and the magnification of
image $i$ is $|\partial\xx_i/\partial\uu|$ \citep[see, e.g.,][]{SEF}.
The second equality uses the fact that $|\partial\xx/\partial\uu|$
is the Jacobian of the transformation between the source and image
planes.  While the sum makes the middle expression look complicated,
it just ensures that the probability density is carried into every
part of the image plane, with no gaps or overlaps.\footnote{The way
to think about this is that the full source plane maps into the
full image plane, and each image position corresponds to a single
source.}  The bottom line is that with $\nu=2$, the magnification
weighting in the source plane corresponds to {\it uniform\/}
weighting in the image plane.  Therefore, what we can do is sample
image positions uniformly within the multiply-imaged region of the
image plane, and then map them back to the source plane to obtain
a set of random sources with magnification weighting.

Apart from ease of use, this transformation has one additional
advantage and one small drawback.  The drawback is that we have
had to assume a source luminosity function of the form
$dN/dS \propto S^{-2}$; but this is not so different from the
luminosity function of sources in the largest existing lens
survey (see \refsec{mock2}).  The practical goal of including
magnification bias is to avoid undersampling quad lenses, and
for this purpose the $\nu=2$ luminosity function is perfectly
adequate.  The fringe benefit of sampling uniformly on the sky
is that each galaxy in the group is automatically weighted by its
lensing cross section, so massive galaxies contribute the most to
our sample of mock lenses just as they would in reality.

In creating the final catalog, choosing the source density
represents a compromise between adequate sampling of the range
of lens properties and the computational time required for the
lens modeling.  We find that a sampling density (in the image
plane) of 40/$\Box\arcsec$ yields a total of 2606 doubles and
1043 quads, which seems sufficiently large but still manageable.
The number of mock lenses associated with each of the 13 galaxies
is given in \reftab{grp}.  Once we have created the mock lenses,
we imagine observing them with the following measurement
uncertainties: $0\farcs003$ in the positions, which is typical
of modern data \citep[from Hubble Space Telescope images or radio
interferometry, see][]{lehar,patnaik,trotter}; 10\% uncertainties
in the fluxes; and time delay uncertainties of 2 days, which is
somewhat better than much current data
\citep[e.g.,][]{1115tdel,burud02a,burud02b} but achievable with
dedicated campaigns \citep[e.g.,][]{1608tdel,colley}.

The mock lenses require that we specify the cosmology and the
redshifts of the source and lens.  We adopt a cosmology with
$\Omega_M=0.3$, $\Omega_\Lambda=0.7$, and $h=0.7$, and we use
the lens redshift $z_l=0.31$ and source redshift $z_s=1.72$ to
mimic PG~1115+080.  Note, however, that these particular values
only affect the time delays.  The time delay between images $i$
and $j$ has the form:
\begin{eqnarray}
  \Delta t_{ij} &=& \frac{1+z_l}{c}\,\frac{D_{ol} D_{os}}{D_{ls}}
    \biggl[ \Bigl(\phi(\xx_i)-\phi(\xx_j)\Bigr) \nonumber\\
  &&\qquad - \frac{1}{2}\left(\defl{i}^2-\defl{j}^2\right) \biggr] .
\end{eqnarray}
The term in square brackets is expressed in angular units, so
the only physical scale (and the only dependence on redshifts
and cosmology) appears in the factors out front.  Changing the
redshifts or the cosmology would rescale all time delays in the
same way, and would not affect our conclusions.

Once we have the mock lenses, we fit them using the two standard
lens models that are usually applied when nothing is known about
environment.  When a new lens is discovered, the first thing people
usually do is fit it with a singular isothermal ellipsoid
(SIE).\footnote{Saha \& Williams (1997, 2004) have introduced
general non-parametric lens models that do not require the
isothermal assumption, or indeed any other explicit assumptions
about the lens galaxy mass distribution.  However, those models
treat the environment as a simple shear, so they are still subject
to all of the systematic effects that we identify.}  If the fit
is poor, the common next step is to add an external shear ---
to model a possible environmental perturbation using just the
$\gamma_{\env}$ term in \refeq{env}.  (The higher-order terms are
dropped for simplicity and because they are thought to be small;
the $\kappa_{\env}$ term is dropped because it leads to the
mass-sheet degeneracy.)  Lenses are occasionally treated with
more complex environment models
\citep[e.g.,][]{kk1115,bf0957,2016mod,kneib}, but those models
are customized to detailed observational data on the environments.
Our goal is to identify the kinds of problems that arise from
standard analyses when little or nothing is known about lens
environments.

\subsection{Application to lens statistics}
\label{sec:mock2}

For studying the statistics of lens samples, it is useful to
make two small modifications to the mock lens catalog.  First,
in this analysis we are not limited by the computational effort
of modeling the lenses; all we want to do is generate the catalog
and then determine its statistical properties.  We can therefore
increase the sampling density to $\sim\!10^{5}/\Box\arcsec$ to
improve the precision of the Monte Carlo calculations.

The second modification relates to magnification bias.  We
would like to mimic the largest existing lens survey, the Cosmic
Lens All-Sky Survey \citep[CLASS;][]{CLASS}.  The luminosity
function of sources in CLASS is well described as a power law
$dN/dS \propto S^{-\nu}$ with $\nu=2.1$ \citep{rusinQD,chae}.
This is sufficiently similar to $\nu=2$ to justify the simplifying
transformation used in \refsec{mock1} (see eq.~\ref{eq:pimg}).
However, since we are modifying the catalog for the statistical
analysis anyway, we go ahead and use $\nu=2.1$ to allow a more
direct comparison with CLASS.\footnote{Using $\nu=2.1$ prevents
us from repeating the coordinate transformation used in
\refsec{mock1}; so we select points uniformly in the source
plane and apply the magnification weighting {\it a posteriori},
as indicated by \refeq{Bsig2}.}

The first statistical step is to sum over the sources associated
with a particular galaxy to obtain that galaxy's lensing cross
section.  In fact, the quantity of interest is the product of
the lensing cross section and the magnification bias, which we
term the ``biased cross section,'' and which can be written as
\begin{equation} \label{eq:Bsig}
  \Bsig = \int_{\rm mult} \frac{N(S_0/\mu(\uu))}{N(S_0)}\ d\uu\,,
\end{equation}
where the integral extends over the multiply-imaged region of
the source plane, and $N(S)$ is again the number of sources
brighter than flux $S$.  For a power law source luminosity
function $dN/dS \propto S^{-\nu}$, \refeq{Bsig} simplifies to
\begin{equation} \label{eq:Bsig2}
  \Bsig = \int_{\rm mult} \mu(\uu)^{\nu-1}\ d\uu\,.
\end{equation}
By integrating over the full multiply-imaged region behind a
galaxy, we obtain the total biased cross section.  With similar
integrals restricted to the doubly-imaged and quadruply-imaged
regions, we can compute the biased cross sections for double
and quad lenses separately.

The next statistical step is to sum over galaxies to compute
the lensing optical depth $\tau$, which represents the total
probability that a given source is lensed.  In general, we
would consider some appropriate population of galaxies described
by a mass function $dn/dM$ \citep[e.g.,][]{TOG,csk93a}:
\begin{equation} \label{eq:tau}
  \tau = \frac{1}{4\pi} \int dV \int dM\ \frac{dn}{dM}\ \Bsig\,,
\end{equation}
where the first integral is over the volume between the observer
and the source.  Replacing $\Bsig$ with the biased cross section
for doubles or quads, we can compute the optical depths $\tau_2$
and $\tau_4$ for 2-image and 4-image lenses.  The predicted ratio
of quad to double lenses is then simply $\tau_4/\tau_2$.  In
practice, we have only 13 discrete galaxies, so we replace the
mass integral $\int (dn/dM)\,dM$ with a sum over the group
galaxies.  Also, we know the group redshift, so we drop the
volume integral.

To understand how the optical depth is used to place constraints
on $\Omega_\Lambda$, consider a flat universe with a non-evolving
population of lens galaxies.  While this simple assumption appears
to be surprisingly good \citep{schade,im,ofek,chae2}, allowing
some evolution would change some quantitative details but not
affect the thrust of our argument \citep[see][]{mitchell}.  In
the non-evolving case the optical depth takes the form
\begin{equation} \label{eq:tau2}
  \tau(z_s) = \Bsig \times \Gamma n_{\rm tot} \times D(z_s)^3 \,,
\end{equation}
where $n_{\rm tot}$ is the integrated comoving number density of
galaxies, $\Gamma$ is a dimensionless number that depends on the
shape of the mass function \citep[see, e.g.,][]{csk93a,mitchell},
and $D(z)$ is the comoving distance to redshift $z$.  In other
words, the optical depth is proportional to the volume of the
universe out to the redshift of the sources.  Conceptually, one
takes a measurement of the optical depth, adds data or models for
the mass function and a model for the cross section, and uses
\refeq{tau2} to place constraints on the volume of the universe
and hence $\Omega_\Lambda$.

\section{Identifying the Problems}
\label{sec:problems}

\subsection{Basic results: $\chi^2$, $e$, and $h$}
\label{sec:basic}

\begin{figure}
\plotone{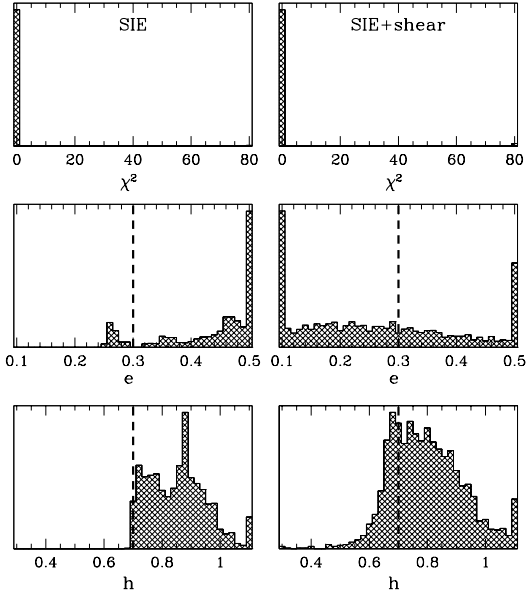}
\caption{
Results for models of mock double lenses.  SIE models are shown in
the left column, and SIE+shear models in the right.  The panels
show histograms of $\chi^2$ values (top), the inferred lens galaxy
ellipticity (middle), and the Hubble constant (bottom).  (Note that
all values outside the $x$-axis range are placed in the left- and
right-most bins.)  The input values $e=0.3$ and $h=0.7$ are indicated
by dashed lines.  
}\label{fig:d-eh}
\end{figure}

We first consider some of the basic quantities derived from
models of individual lenses: the goodness of fit $\chi^2$, the
mass ellipticity $e$, and the Hubble parameter $h$.  \reffig{d-eh}
shows histograms of these quantities for SIE and SIE+shear models
of the mock double lenses in the catalog from \refsec{mock1}.
Doubles provide just 7 observables: position and flux for each
image, and the time delay.  SIE models have 7 parameters: $b$,
$e$, and $\theta_0$ for the galaxy, position and flux for the
source, and $h$.  SIE models of doubles therefore have
$N_{\rm dof}=0$, and in general it is possible to find models that
fit the data perfectly ($\chi^2=0$).  Unfortunately, there are
large errors in the recovered mass ellipticity and Hubble parameter
caused by the neglect of environment.  The problem is worse with
SIE+shear models; even though environment is included (in the
shear approximation), it is completely unconstrained and so there
are enormous uncertainties in the models.  {\it Doubles suffer a
fundamental problem: with so few constraints, lens models cannot
even detect an environmental component in the lens potential, much
less constrain it well enough to yield accurate results.\/}

\begin{figure}
\plotone{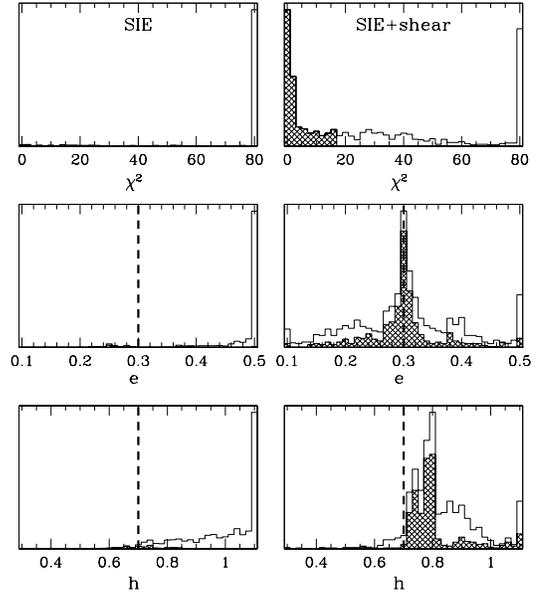}
\caption{
Similar to \reffig{d-eh}, but for mock quad lenses.  The open
histograms show all quads, while the shaded histograms show only
those where the lens model provides an acceptable fit to the data
(models that are not ruled out at the 99\% confidence level; see
text).
}\label{fig:q-eh}
\end{figure}

\reffig{q-eh} shows that the situation is better with quad lenses
because the additional images provide more constraints.  Quads
offer 15 observables: position and flux for each image, and three
independent time delays.  So both SIE models (7 parameters) and
SIE+shear models (9 parameters) are well constrained.  In fact,
SIE models generally give terrible fits: among our mock quads the
median $\chi^2$ is 464, and in 95\% of the lenses the model can
be ruled out at more than 99\% confidence ($\chi^2>20.1$ for
$N_{\rm dof}=8$).  In other words, quads have enough constraints
to make it obvious that environment cannot be neglected.  In this
case the significant errors in $e$ and $h$ are irrelevant because
the $\chi^2$ already reveals that the models are wrong.  Like real
lenses \citep{kks}, our mock lenses indicate that quads cannot be
well fit by models that neglect environment.

The SIE+shear models of quads are more interesting.  Now the median
$\chi^2$ among the mock quads is 19.3, and in only 52\% of the
systems can the model be excluded at 99\% confidence ($\chi^2>16.8$
for $N_{\rm dof}=6$).  In the other 48\% the SIE+shear model provides
an acceptable fit to the data.  Thus, approximating the environment
as a shear is sufficient to provide a good fit to many quads.
{\it This approximation is not adequate, however, for recovering
accurate values for the ellipticity and Hubble parameter.\/}  The
errors in these quantities, while smaller than for doubles, are
still worrisome.

Particularly troubling is the fact that the errors are not random
but biased.  For doubles, SIE models usually overestimate both
$e$ and $h$: the median recovered ellipticity is $e=0.47$; and
$(99,66,30,7)\%$ of the mock doubles yield $h>(0.7,0.8,0.9,1.0)$.
For quads, SIE+shear models do reasonably well with ellipticity:
88\% of the quads with acceptable fits (and 74\% of all quads)
have ellipticity errors $|\Delta e| < 0.1$.  But $h$ is still a
problem, as $(98,26,14,7)\%$ of the quads with acceptable fits
yield $h>(0.7,0.8,0.9,1.0)$.  Poor knowledge of environment can
clearly cause not just uncertainties but also significant biases
in lensing measurements of ellipticity and $H_0$.

In the remainder of the paper we use the SIE lens models for
double lenses (because they give perfect fits, and SIE+shear
models are underconstrained), and the SIE+shear models for quads.  

\subsection{Mass and velocity dispersion of the lens galaxy}
\label{sec:mass}

It is often said that lensing robustly measures the mass of the
lens galaxy within the Einstein radius.  To be precise, this
statement applies to quad and Einstein ring lenses, while for
doubles lensing measures a mass--radius relation of the form
$M(R_1)/R_1+M(R_2)/R_2 = \pi\,\Sigma_{\rm crit}(R_1+R_2)$ where
$R_1$ and $R_2$ are the image radii and $M(R)$ is the aperture
mass (see \citealt{rkk} for a detailed discussion).  (Quads
and rings can actually be thought of in the same way, but with
$R_1=R_2=\Rein$ since all the images appear near the Einstein
radius.)  For simplicity we limit our discussion to the mass
within the Einstein radius as appropriate for quads and rings,
keeping in mind that it would be modified slightly for doubles.

An important caveat is that lensing measures only the total
projected mass within the Einstein radius.  The total certainly
includes contributions from the lens galaxy (its stellar component
and dark matter halo).  But there is also a contribution of
$\pi\,\Rein^2\,\kappa_{\env}\,\Sigma_{\rm crit}$ from the
convergence.  Physically, this mass might represent a smooth
background in which the galaxy is embedded, such as a common group
dark matter halo; or it might represent mass in the immediate
foreground or background, such as the halos of other group member
galaxies that overlap the line of sight.  Models that neglect
convergence assume that all of the measured mass must come from
the lens galaxy, and hence overestimate the galaxy mass.

\begin{figure}
\plotone{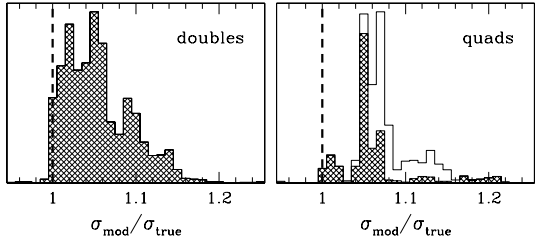}
\caption{
Errors in the velocity dispersion of the lens galaxy;
$\sigma_{\rm true}$ is the true (input) value, while
$\sigma_{\rm mod}$ is the value inferred from lens models.  The
left panel shows doubles; the right panel shows quads, where the
open histogram show all quads while the shaded histograms show
only those where the model fits the data.  The dashed lines shows
where the ratio is unity.
}\label{fig:hist-sig}
\end{figure}

To examine this effect, it is more instructive to quote the
velocity dispersion of the lens galaxy; while the mass depends
strongly on the aperture and is not directly observable, the
velocity dispersion depends only weakly on aperture (not at all
for isothermal models) and is observable (actually the velocity
dispersion of the stars; see below).  For isothermal models, the
mass within the Einstein radius is
$M \propto b\,\Rein \propto \sigma^2\,\Rein$ (see eq.~\ref{eq:b}),
so the conversion is simple.  \reffig{hist-sig} shows that the
models do slightly overestimate the velocity dispersion, with a
typical error of $\sim$6\% (and a broad range, especially for
doubles).  While small, the errors are larger than the typical
$\sim$1\% measurement uncertainties \citep[see][]{rkk}, and they
represent a systematic shift rather than random errors.

The errors may bear on a historic controversy in lensing analyses:
the relation between the velocity dispersion of the dark matter
(the $\sigma$ parameter in isothermal models) and that of the
stars.  \citet{gott} argued that the ratio should be
$\sigma_{\rm DM}/\sigma_{\rm stars} = (3/2)^{1/2}$ if galaxy
luminosity densities are $r^{-3}$ power laws, but studies of more
realistic models for the luminosity indicated that the
{\it central\/} stellar velocity dispersion $\sigma_0$ actually
obeys $\sigma_{\rm DM}/\sigma_0 \approx 1$
\citep{franx,csk93b,csk94}.  The difference may seem small in
the velocity dispersion, but it looms large in lens statistics
because the lensing optical depth scales as
$\tau \propto \sigma_{\rm DM}^4$.  Most recently, \citet{tk04}
have found $\langle\sigma_{\rm DM}/\sigma_0\rangle = 1.15\pm0.05$
by combining direct measurements of $\sigma_0$ for five lens galaxies
with model determinations of $\sigma_{\rm DM}$.  We cannot measure
this ratio in our simulations since we do not include a stellar
component.  However, we can say that if standard models are
overestimating $\sigma_{\rm DM}$ then this effect could explain
part but not all of the difference of the measured value from
unity.

\subsection{Image magnifications}
\label{sec:mag}

In some applications lensing is used as a natural telescope that
provides high resolution for studying the structure of quasars
and their host galaxies \citep[e.g.,][]{rix01,kkm,peng,richards}.  
In this case accurate lens models are essential for ``de-lensing''
the system, or removing the effects of lensing magnification and
distortion to infer the intrinsic properties of the source.  In
other applications the need to know the magnification and distortion
is less obvious but no less important (see \refsec{substr}).  We
consider how reliable lens models are for such applications,
focusing on the magnifications of unresolved images since we use
only point-like sources.

When studying magnifications it is important to separate images
of different types: images that lie at minima of the time delay
surface and have positive parity, as opposed to images that lie
at saddlepoints of the time delay surface and are parity-reversed
\citep[see, e.g.,][]{SEF}.  A double has one minimum and one
saddlepoint, while a quad has two of each.  The image type has
become particularly important in studies of microlensing and
millilensing (see \refsec{substr}).  If the lensed images are
point-like then the parities cannot be observed directly, but
image parities can almost always be determined unambiguously with
even simple lens models \citep[e.g.,][]{saha03}.

\begin{figure}
\plotone{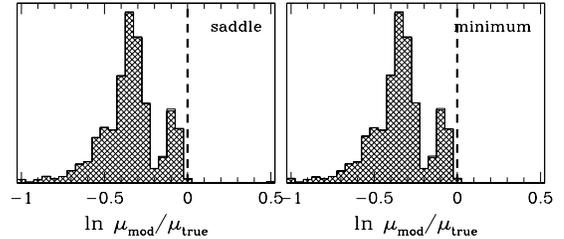}
\caption{
Histograms of the ratio of the lensing magnification inferred from
lens models to the true magnification, for mock double lenses.  The
panels refer to the two image classes (saddlepoints and minima).
The dashed lines show where the ratio is unity.
}\label{fig:d-mu}
\end{figure}

\reffigs{d-mu}{q-mu} show histograms of the error in the lensing
magnification, for doubles and quads respectively.  {\it We see
that standard lens models tend to underestimate the magnifications,
often by a factor of $\sim$1.5--2 and sometimes much more.\/}
The important qualitative point is that the errors are biased:
the models almost always underestimate the magnifications.

\begin{figure}
\plotone{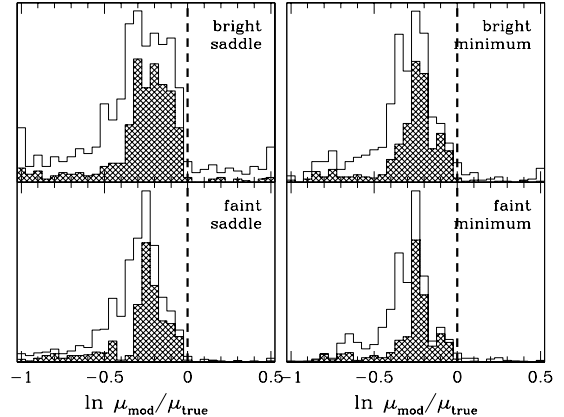}
\caption{
Similar to \reffig{d-mu}, but for mock quad lenses.  The panels
refer to the four image classes.  The open histograms show all
quads, while the shaded histograms show only those where the lens
model provides an acceptable fit to the data.
}\label{fig:q-mu}
\end{figure}

\subsection{Constraints on small-scale structure}
\label{sec:substr}

Lensing offers unique constraints on small-scale structure in
lens galaxies via the phenomena of microlensing by stars
\citep[e.g.,][]{chang,paczynski,schmidt,wyithe} and millilensing
by dark matter clumps \citep[e.g.,][]{mao,metcalf,chiba,dk}.  The
question is whether environment-related errors in lens models can
affect the results.  The millilensing constraints, in particular,
are derived from lenses with ``flux ratio anomalies,'' or flux
ratios that are inconsistent with smooth lens potentials.
Conceptually, the probability of having a flux ratio anomaly is
given by the abundance of mass clumps times the cross section for
any given clump to cause millilensing (to significantly perturb
the flux of an image).  Converting the abundance of flux ratio
anomalies to a clump abundance therefore requires the ability
to compute a clump's millilensing cross section.  But that cross
section depends in a complicated way on the total lens potential.
The dependence is not apparent in the detailed simulations used
by \citet{dk} and \citet{metcalf2237} to derive clump abundances,
but it is made explicit in the analytic expressions derived by
\citet{clump} for the millilensing cross sections of clumps modeled
as isothermal spheres.

\begin{figure}
\plotone{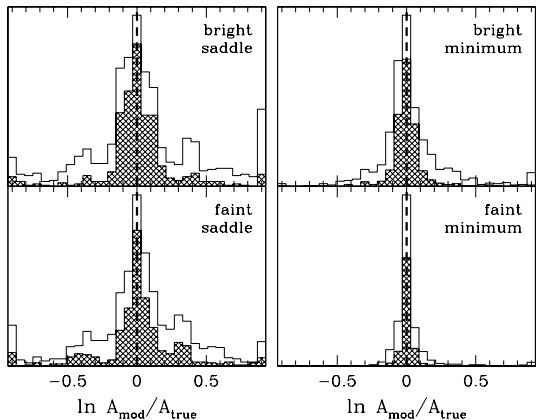}
\caption{
Errors in the estimated cross section for millilensing by a mass
clump in front of an image in a mock quad lens.  The histograms
show the ratio of the cross section computed with the fitted lens
model, to that computed with the true lens potential.  The dashed
lines show where the ratio is unity.  The open histograms show all
quads, while the shaded histograms show only those where the lens
model provides an acceptable fit to the data.
}\label{fig:q-sub}
\end{figure}

We focus on millilensing because there are as yet no simple
expressions for microlensing cross sections.  We compute the cross
section for an isothermal clump to produce a 30\% change in the
flux of an image.  \reffig{q-sub} shows the ratio of the cross
section computed with an SIE+shear lens model to the cross section
computed with the true lens potential, for the mock quad
lenses.\footnote{Small-scale structure analyses are usually
restricted to quads, because doubles have too few constraints to
allow robust identification of flux ratio anomalies.}  Studying
the four types of images separately is important because saddlepoint
images --- especially the bright saddles --- are believed to be
particularly susceptible to perturbations; this effect provides
a unique signature of lensing by small-scale structure that allows
it to be distinguished from other effects \citep{sw,kd}.

There are indeed errors in the millilensing cross sections due to
errors in the lens models.  Interestingly, the error distributions
appear to be symmetric in $\ln(A_{\rm mod}/A_{\rm true})$.  They
tend to have relatively tight cores but broad wings, so the RMS
error in $\ln(A_{\rm mod}/A_{\rm true})$ is 0.29 for the bright
saddles, 0.31 for the faint saddles, 0.14 for the bright minima,
and 0.09 for the faint minima.  (We are considering only quads
where the SIE+shear model provides an acceptable fit.)  The fact
that bright saddlepoint images are so sensitive to the model errors
is troubling because they are so important for millilensing analyses.

One would naively expect that a $\sim$30\% uncertainty in the
millilensing cross section might cause a comparable uncertainty in
the inferred clump abundance, which would make this effect smaller
than other statistical and systematic uncertainties in current
results \citep[see][]{dk}.  However, to make that statement
quantitative and precise we would need to replicate complete
millilensing analyses \citep[following, e.g.,][]{dk,metcalf2237}.
As for microlensing, without a simple way to compute perturbation
cross sections we can only guess that there would be qualitatively
similar results.  Clearly there is much more to do here; we just
want to point out that analyses of small-scale structure depend
on the overall lens model, and errors in that model related to
environment cannot be ignored.

\subsection{Constraints on $\Omega_\Lambda$}
\label{sec:lambda}

\begin{figure}
\plotone{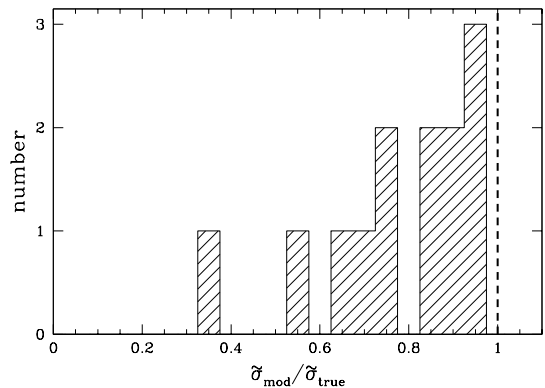}
\caption{
Histogram of errors in the lensing cross section; $\Bsig_{\rm mod}$
is the biased cross section with the galaxy assumed to be isolated,
while $\Bsig_{\rm true}$ is the value with the galaxy in its proper
environment.
}\label{fig:csec-hist}
\end{figure}

We now switch gears and turn to statistical analyses of lens
populations and the constraints they yield on the cosmological
constant $\Omega_\Lambda$.  As discussed in \refsec{mock2} (see
eq.~\ref{eq:tau2}), lensing limits on $\Omega_\Lambda$ depend
on the optical depth $\tau$, so the key question is whether
environment causes errors in $\tau$.  First, though, it is useful
to ask whether there are errors in the lensing cross section
(from which the optical depth is determined).  To answer this
question, we compare the biased cross section $\Bsig_{\rm true}$
that a galaxy has when placed in its proper environment to the
biased cross section $\Bsig_{\rm mod}$ the {\it same\/} galaxy
would have if it were assumed to be isolated.  Any differences
indicate that neglecting environment does cause errors in the
cross section.  \reffig{csec-hist} shows the errors in the form
of a histogram of the ratio $\Bsig_{\rm mod}/\Bsig_{\rm true}$.
{\it Models that neglect environment underestimate the cross
section by tens of percent or more.\/}  The underestimate is
caused entirely by the magnification bias,\footnote{Curiously,
the radial caustic of an isothermal lens is not affected by
convergence and shear from the environment; the radial caustic
is the set of points that map to the origin, and the deflection
from convergence and shear vanishes at the origin (see
eq.~\ref{eq:env}).  Except in the rare case of naked cusps, the
unbiased cross section is simply the area enclosed by the radial
caustic, so it is insensitive to convergence and shear.} and
arises because standard models underestimate the magnifications
(see \refsec{mag}).

\begin{figure}
\plotone{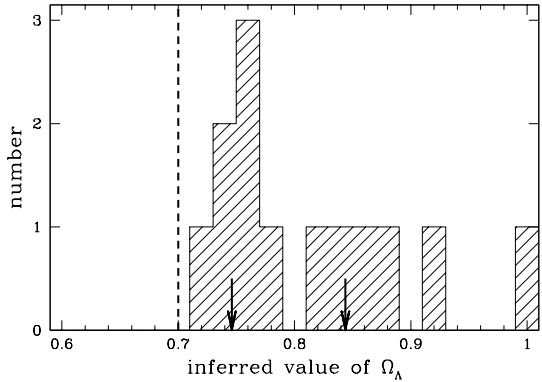}
\caption{
Errors in the inferred value of $\Omega_\Lambda$, assuming a
typical source redshift $z_s \sim 1.27$ as appropriate for CLASS
\citep{marlow,chae}.  The histogram shows the values for the 13
galaxies taken separately.  The dashed line shows the input value
$\Omega_\Lambda=0.7$.  The arrow at $\Omega_\Lambda=0.84$ shows
the value if we average over the 13 galaxies (weighting each by
its cross section).  The arrow at $\Omega_\Lambda=0.75$ shows the
value if we assume that 25\% of all lens galaxies lie in groups
like our mock group, while the other 75\% are isolated.
}\label{fig:lam-hist}
\end{figure}

The next question is how errors in the cross section propagate
into errors in $\Omega_\Lambda$.  \reffig{lam-hist} gives a
sense of the effect, by showing the error in $\Omega_\Lambda$
that would result from each of the cross section errors in
\reffig{csec-hist}.  {\it Underestimates in the cross sections
lead directly to overestimates in $\Omega_\Lambda$, which can
be quite large.\/} This figure is a bit unfair, because
constraints on $\Omega_\Lambda$ properly depend on the optical
depth, and not all galaxies contribute equally.  For example,
the point at $\Omega_\Lambda \approx 1$ (corresponding to the
point with $\Bsig_{\rm mod}/\Bsig_{\rm true}=0.36$ in
\reffig{csec-hist}) is produced by galaxy \#5, which is strongly
perturbed by galaxy \#3.  However, galaxy \#5 has a small cross
section and contributes little to any lens sample, so its
properties are not very important.  A better approach is to
compute the optical depth by summing the cross sections, and
then compare the ``model'' (galaxies assumed to be isolated) and
true optical depths.  This lead to an error
$\tau_{\rm mod}/\tau_{\rm true}=0.70$, and hence an inferred
value $\Omega_\Lambda=0.84$ (indicated by an arrow in
\reffig{lam-hist}).

This result holds if all lens galaxies lie in environments
like our mock group, which is unlikely.  The final effect of
environment on lens statistics cannot be determined because the
distribution of lens galaxy environments is unknown.  Still, we
can make a crude estimate.  \citet{kcz} use galaxy demographics
to predict that at least $\sim$25\% of lens galaxies lie in
groups or clusters.  If we conservatively assume that the fraction
of lenses with groups is indeed 25\% while the other 75\% are
isolated, we would find a cross section error
$\tau_{\rm mod}/\tau_{\rm true}=0.90$ and hence an inferred value
$\Omega_\Lambda=0.75$ (also indicated in \reffig{lam-hist}).  Even
this simple and conservative estimate points to the importance of
measuring lens galaxy environments and understanding their effects
on lens statistics.

\subsection{The quad/double ratio}
\label{sec:quaddoub}

A long-standing puzzle in lens statistics is why so many lenses
are quads rather than doubles \citep{king,csk96b,kks,rusinQD,cohn}.
In the CLASS statistical sample of 13 lenses, 7 are doubles and 5
are quads \citep{CLASS}.  (The remaining lens is a complicated
system that has six images because there are three lens galaxies;
see \citealt{1359}.)  The quad/double ratio depends on the relative
cross sections for 4-image and 2-image lenses (along with
magnification bias), and it is thought to depend mainly on the
ellipticity of lens galaxies.  An observed quad/double ratio near
unity is said to require a typical ellipticity $e \sim 0.6$ that
is much larger than observed or predicted for normal galaxies
\citep{csk96b,kks,rusinQD}.  In fact, some have spoken of the
high observed ratio creating an ``ellipticity crisis'' in lensing
\citep[see][]{csk96b}.

Conventional wisdom holds that environment has little effect on
the quad/double ratio, because (when averaged over orientation)
shear does not significantly change the 2-image and 4-image cross
sections \citep[e.g.,][]{rusinQD}.  However, \citet{cohn} recently
showed that low-mass satellites around lens galaxies, which are
common and not very sensitive to the larger environment, can roughly
double the predicted quad/double ratio.  We can now determine the
effects of a larger group environment, using a model that is more
realistic than simple shear.

\begin{figure}
\plotone{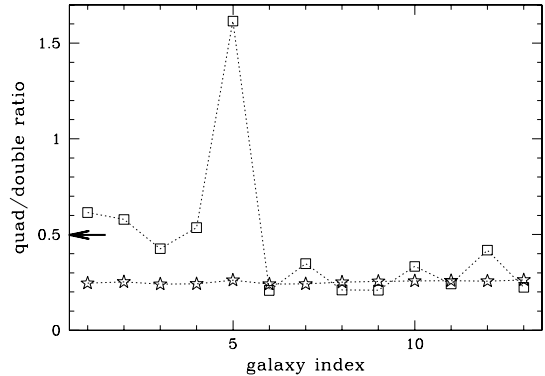}
\caption{
Quad/double ratios for the 13 galaxies, computed as the ratio of
the biased cross section for 4-image lenses to that for 2-image
lenses.  The stars show the results with each galaxy assumed to
be isolated; the galaxies should have the same quad/double ratio
since they have the same ellipticity, so the fluctuations indicate
the level of numerical noise ($\sim$3\%).  The squares show the
results with each galaxy placed in its proper environment.  (The
lines are drawn just to guide the eye.) The arrow at $Q/D=0.50$
shows the net ratio when all the galaxy cross sections are summed.
}\label{fig:quaddoub}
\end{figure}

Under the standard assumption that galaxies are isolated, we would
compute a quad/double ratio of 0.25 for each of our galaxies (the
same because they all have the same ellipticity).  By contrast,
placing the galaxies in their proper environments yields the
quad/double ratios shown in \reffig{quaddoub}.  We find that
neglecting environment causes significant errors in the quad/double
ratio.  More relevant than the individual ratios is the net ratio
after summing the 2-image and 4-image cross sections for all the
galaxies; this represents the ratio that would be expected for a
real lens survey.  The net quad/double ratio is again 0.25 when
environment is ignored, but 0.50 when environment is included.
{\it In other words, neglecting environment causes a significant
underestimate of the quad/double ratio.\/}

Note that our predicted quad/double ratio of 0.50 is higher than
the ratio $1043/2606=0.40$ in our mock lenses.  Part of the
difference may be due to statistical fluctuations in the mock lens
catalog, because some of the galaxies produce a small number of
lenses (see \reftab{grp}).  However, most of the difference arises
from magnification bias.  The mock lenses were generated using
magnification bias corresponding to a source luminosity function
$dN/dS \propto S^{-\nu}$ with $\nu=2$ (see \refsec{mocklens}).  By
contrast, the lens statistics calculations use a slightly steeper
source luminosity function with $\nu=-2.1$ (as appropriate for
comparison to CLASS).  The steeper luminosity function produces
a larger magnification bias that increases the weight of quads
relative to doubles.

Even with environment included our predicted quad/double ratio is
still somewhat smaller than observed.  Part of the explanation
may be that we have assumed a particular ellipticity, $e=0.3$.
Increasing the ellipticity to $e=0.4$ but holding everything else
fixed would yield a quad/double ratio of 0.67 with environment
included, which is close to the observed ratio.  (With $e=0.4$
and environment omitted, the quad/double ratio would be 0.40.)
A proper analysis would have to include an appropriate distribution
of ellipticities.  Another part of the explanation may be that we
have considered just one mock group, and the full distribution of
lens environments will need to be taken into account.  A third
issue is that we have not considered low-mass satellite galaxies,
which would further increase the quad/double ratio \citep{cohn}.
Still a fourth possibility is that the high observed quad/double
ratio is a statistical fluke.  Thorough study of these possibilities
is beyond the scope of this paper.  We mainly want to point out
that environment represents a systematic effect that can
significantly increase the predicted quad/double ratio without
violating other constraints \citep[c.f.][]{rusinQD}.

\section{Fixing the Problems}
\label{sec:fix}

Having identified some of the environment-related problems in
lensing analyses, we briefly consider what causes them and how
they can be fixed.  (A fuller treatment of these issues will be
given in the follow-up paper.)  For doubles the situation is clear:
with so few constraints, lens models are unable to recognize and
constrain an environmental contribution to the lens potential;
to fix the errors, it will be necessary to study the environment
separately and impose it on the lens models.

\begin{figure}
\plotone{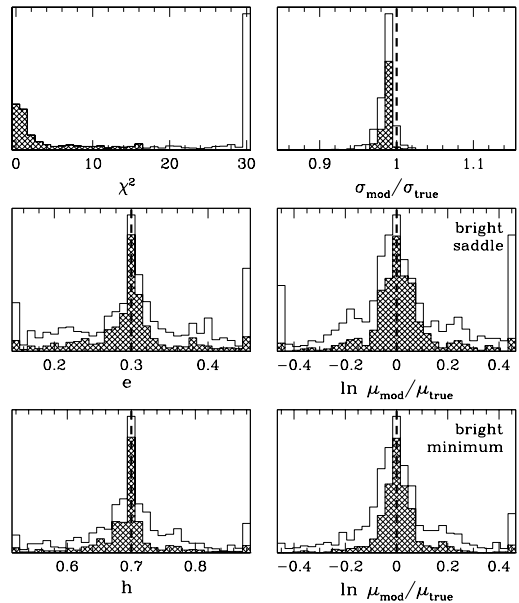}
\caption{
Sample results for mock quad lenses, for SIE+shear models that
also include the convergence $\kappa_{\env}$ from the environment.
We show histograms of the goodness of fit $\chi^2$, the ellipticity,
and the Hubble parameter (left), as well as errors in the lens
galaxy velocity dispersion and the magnifications of the bright
saddle and bright minimum images (right); this figure is to be
compared with Figures~\ref{fig:q-eh}, \ref{fig:hist-sig}, and
\ref{fig:q-mu} (but note the different axis scales).  As before,
the open histograms show all quads while the shaded histograms
show those where the model provides an acceptable fit.
}\label{fig:grp1}
\end{figure}

For quads, lens models are often able to constrain shear from the
environment, but the shear approximation clearly fails to capture
all of the environmental effects.  Returning to \refeq{env}, the
question is whether the errors are caused mainly by neglect of the
convergence term ($\kappa_{\env}$, representing the additional
mass at the position of the lens galaxy), or by neglect of the
higher-order terms.  To address this question, we consider what
would happen if we somehow knew $\kappa_{\env}$ for each galaxy
and could include it in the models.  According to the mass-sheet
degeneracy \citep{gorenstein,saha}, the $\kappa_{\env}$ term
simply rescales some of the model quantities without affecting
the goodness of fit; some of the key rescalings are:
\begin{eqnarray}
  b      &\propto& (1-\kappa_{\env})\,, \\
  \sigma &\propto& (1-\kappa_{\env})^{1/2}\,, \\
  h      &\propto& (1-\kappa_{\env})\,, \\
  \mu    &\propto& (1-\kappa_{\env})^{-2}\,, \\
  \Bsig  &\propto& (1-\kappa_{\env})^{-2(\nu-1)}\,,
\end{eqnarray}
where $\sigma$ is the inferred velocity dispersion while $\Bsig$
is the biased cross section, and $\nu$ is the power law index of
the source number counts (see \refsec{mocklens}).
\reffigs{grp1}{lam-hist2} show that this rescaling makes the
model results much more accurate.  In particular, it removes the
{\it biases\/} in quantities such as the ellipticity, Hubble
parameter, lens galaxy velocity dispersion, image magnifications,
and $\Omega_\Lambda$.  There are still some random errors: 0.07
in $e$ and $h$; and 0.20 and 0.16 in $\ln\mu$ (for the bright
saddle and bright minimum images, respectively).\footnote{These
are the numbers for quads in which the lens models provide an
acceptable fit (the shaded histograms in \reffig{grp1}).}
Interestingly, adding the convergence does not fix the
underestimated quad/double ratio, because it rescales the 2-image
and 4-image cross sections in the same way leaving the ratio
unchanged.  {\it Still, it appears that neglecting the convergence
from the environment is the primary cause of the biases in lensing
results for quads.\/} Whether the residual scatter is due to neglect
of the higher-order terms in \refeq{env} or just to observational
noise is not clear from this analysis.

\begin{figure}
\plotone{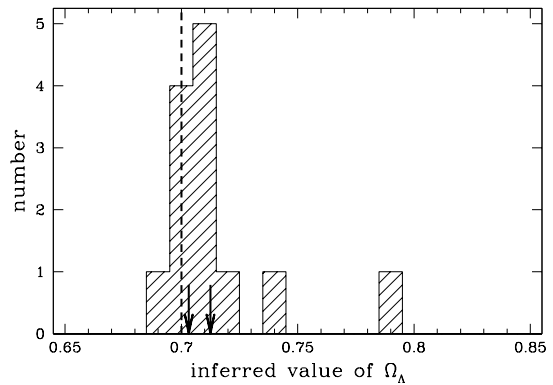}
\caption{
Similar to \reffig{lam-hist}, but for SIE+shear models that also
include the convergence $\kappa_{\env}$ from the environment.
The dashed line shows the input value $\Omega_\Lambda=0.7$.  The
arrow at $\Omega_\Lambda=0.712$ shows the value if we average
over the 13 galaxies (weighting each by its cross section).  The
arrow at $\Omega_\Lambda=0.703$ shows the value if we assume that
25\% of all lens galaxies lie in groups like our mock group, while
the other 75\% are isolated.
}\label{fig:lam-hist2}
\end{figure}

The problem is that the convergence is not directly observable,
because it usually represents dark matter.  Nor is there a reliable
way to estimate the convergence from properties of lens models.
This statement may seem surprising, because for simple environments
dominated by a single spherical halo there is a general relation
between shear --- which can be inferred from lens models, at least
for quads --- and convergence \citep{miralda-escude,kaiser}:
\begin{equation}
  \gamma_{\env} = \langle\kappa_{\env}\rangle_{R} - \kappa_{\env}\,,
\end{equation}
where $R$ is the offset between the lens galaxy and the
environment halo, and $\langle\kappa_{\env}\rangle_{R}$ is the
mean convergence (or surface mass density in units of the critical
density for lensing) in an aperture of radius $R$ centered on the
environment halo.  Because of the average, $\kappa_{\env}$ and
$\gamma_{\env}$ depend on radius and on the halo properties (mass,
concentration, etc.) in different ways, and there is no universal
relation between them.  A second and more troubling problem is
that the convergence is a scalar while the shear is a rank-2
traceless tensor (or a headless vector with an amplitude and a
direction that is invariant under $180^\circ$ rotations; see
eq.~\ref{eq:env}).  This means that in an environment comprising
multiple mass components, the convergence and shear from the
different components sum in different ways, and attempts to
relate the net convergence to the net shear break down.  A final
problem is that this could never work for doubles anyway, since
there are no good model constraints on the shear.

The only solution is to build a model that allows us to
translate observable quantities into reliable inferences about
the convergence.  If we can identify and model all of the mass
components in the lens environment, then the shear, convergence,
and higher-order terms will all be included self-consistently.
Creating such a model is not as crazy as it may sound, because
many of the key parameters are observable.  For example, the
relative positions of the group member galaxies can be measured,
and the relative magnitudes can be used to derive the $b$ ratios
(recall the Faber-Jackson relation, eq.~\ref{eq:brat}).  Although
the mass ellipticities cannot be directly observed, we can
hypothesize that they are not critical; we can set them to zero
when modeling the mock lenses and see whether we still obtain
accurate model results.  (To be precise, we let the main lens
galaxy be elliptical but make all the {\it other\/} galaxies
circular.)  In other words, all of the new parameters in the most
basic {\it realistic\/} model of environment are either constrained
by observations or fixed, so adding complexity has not caused an
explosion of free parameters.  We now consider whether such models
are sufficient to fix the problems in lensing analyses.  To allow
for realistic noise in the observational constraints on the new
parameters, we include generous uncertainties: measurement errors
of $1\arcsec$ in the relative positions; and intrinsic scatter of
0.08~dex in the $\sigma(L)$ relation \citep{sheth}, or 0.16~dex
in the model $b$ ratios.

\begin{figure}
\plotone{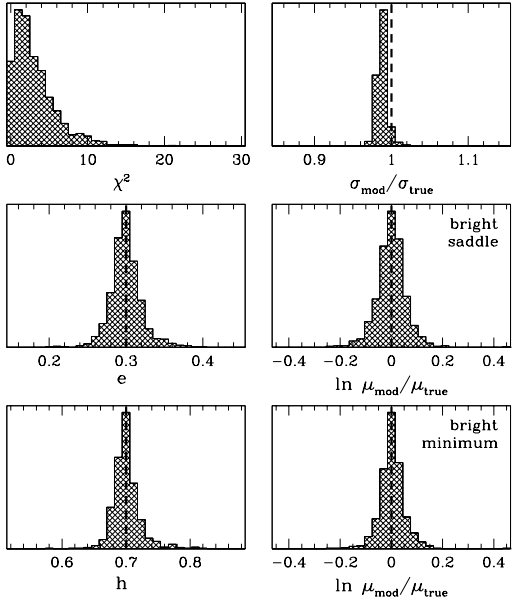}
\caption{
Similar to \reffig{grp1}, but for models that incorporate all of
the group member galaxies using reasonable observational
constraints.  Here all of the quads are well fit by the models.
}\label{fig:q-grp3}
\end{figure}

\reffig{q-grp3} shows the results when we apply such models to
quad lenses.  There are no biases in the results, and the scatter
is encouragingly small: just 0.02 in both $e$ and $h$; and
0.05--0.06 in $\ln\mu$.  In other words, models that include the
environment break the mass-sheet degeneracy and fix the problems
with standard lensing analyses.  Furthermore, these models do
notably better than the models in \reffigs{grp1}{lam-hist2} that
just included convergence, indicating that the higher-order terms
in the lens potential from the environment (eq.~\ref{eq:env}) are
in fact significant.  Note that neither the assumption of circular
group galaxies nor the inclusion of noise causes significant
problems in the models.

\begin{figure}
\plotone{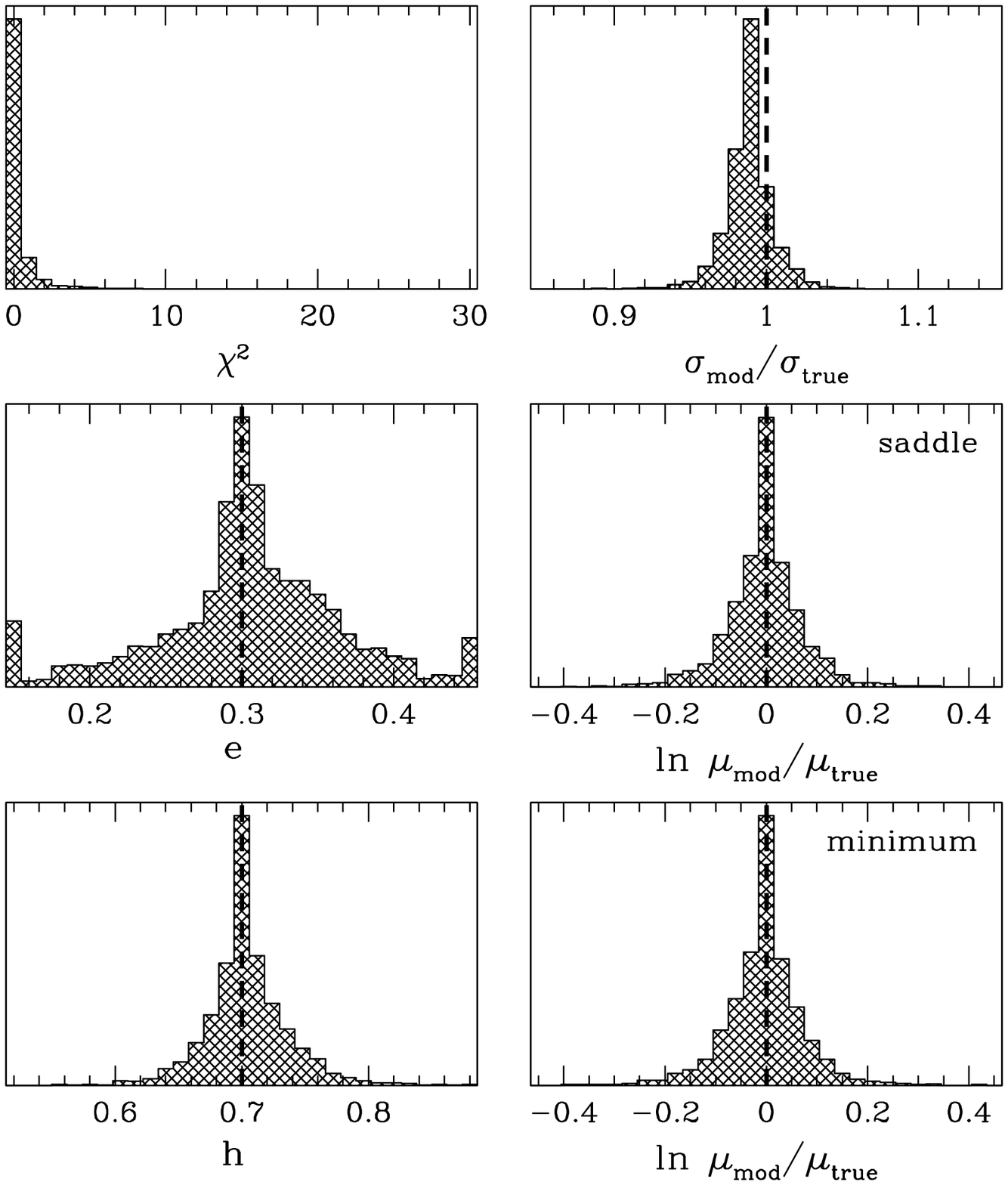}
\caption{
Similar to \reffig{q-grp3}, but for mock double lenses.
}\label{fig:d-grp3}
\end{figure}

\begin{figure}
\plotone{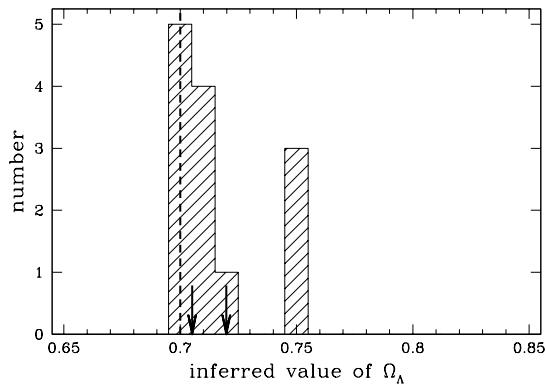}
\caption{
Similar to \reffig{lam-hist2}, but for models that incorporate the
group member galaxies using reasonable observational constraints.
The arrow at $\Omega_\Lambda=0.720$ shows the value if we average
over the 13 galaxies (weighting each by its cross section).  The
arrow at $\Omega_\Lambda=0.705$ shows the value if we assume that
25\% of all lens galaxies lie in groups like our mock group, while
the other 75\% are isolated.
}\label{fig:lam-hist3}
\end{figure}

\reffig{d-grp3} shows that similar results hold for doubles,
with slightly larger scatter: 0.07 in $e$; 0.03 in $h$; and 0.08
in $\ln\mu$.  While we saw in \refsec{problems} that doubles are
largely useless for astrophysical measurements when environments
are unknown, we see now that observing and modeling the
environment makes doubles almost as valuable as quads.  Finally,
\reffigs{lam-hist3}{quaddoub2} show that $\Omega_\Lambda$ and the
quad/double ratio can also be recovered well from statistical models
that account for environment.

\begin{figure}
\plotone{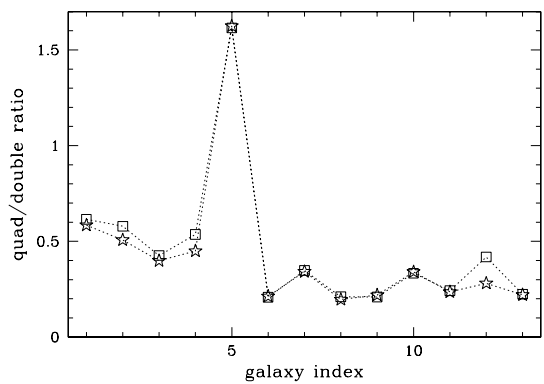}
\caption{
Similar to \reffig{quaddoub}, but for models that incorporate the
group member galaxies.  The squares again show the quad/double
ratios when the galaxies are placed in their true environments,
while the stars show the values using the model environments.  The
net ratio with the model environments is 0.45 compared with the
true value of 0.50.
}\label{fig:quaddoub2}
\end{figure}

\section{Discussion}
\label{sec:disc}

In this paper we have effectively studied 13 different
galaxy/environment configurations, but they were all drawn from
a single mock group of galaxies.  We must therefore acknowledge
the following caveats and questions about our results.

{\it Is our mock group realistic?\/}
We constructed the system to mimic the group around the observed
4-image lens PG~1115+080, and to be similar to nearby X-ray
luminous groups, so it has reasonable properties.  Furthermore,
the distribution of lensing shears produced by the system is
similar to what is measured and predicted for (quad) lenses
\citep[c.f.][]{holder}, so the group's contribution to the lens
potential seems realistic.  However, the proper way to answer
this question is to study more real groups around real lenses in
detail.

{\it Is this group typical?\/}
Our group represents just one sample system.  Based on current
knowledge of lens environments (limited though it is) and on the
shear distribution, we believe that it captures features that
are representative.  But we must identify the full distribution
of lens environments before we can say for sure.

{\it How do the results depend on the group properties?\/}
It seems likely that the systematic effects will scale with the
mass or richness of the environment.  However, that glib
generalization obscures a lot of interesting details.  A group's
contribution to the lens potential may depend on many properties
beyond its total mass: the ellipticities of the group member
galaxies (addressed briefly in this paper); the properties of the
galaxy population, such as the elliptical/spiral and dwarf/giant
ratios; the fraction of mass in the common halo versus that bound
to the galaxies (e.g., the degree to which the galaxy halos have
been truncated by tidal stripping); the radial profile and angular
structure of the common halo; and the offset between the group
centroid and the lens galaxy.  In the limit that the environment
is dominated by a common dark halo, we must also understand how
well observables such as the group centroid and velocity dispersion
will be able to constrain the important properties of the
environment.  Studying these issues in detail is the subject of
the follow-up paper.

Much more detailed understanding of lens environments is clearly
required.  We need to characterize the distribution of lens
environments in order to guide theoretical calculations.  For
lenses lying in groups (or clusters), we must find the member
galaxies and measure the centroid and velocity dispersion of the
system, and then build sophisticated lens models that finally
treat environments properly.  Even for lenses that do not lie in
groups or clusters, careful observations will be needed in order
to establish that the environments are unimportant.

\section{Conclusions}
\label{sec:concl}

Poor groups of galaxies around strong gravitational lens systems
can affect lensing constraints on the masses and shapes of galaxy
dark matter halos, the amount of substructure in dark matter halos,
the quad/double ratio, the properties of lensed sources, the Hubble
constant, and the dark energy density.  Not knowing that a lens
lies in a group can cause biases and uncertainties in lensing
analyses.  Models of double lenses that neglect the environment
will generally fit the data well ($\chi^2 \approx 0$) but yield
parameter values that are grossly incorrect.  Models of quad
lenses will reveal that environment cannot be ignored but will not
fully constrain the environmental component of the lens potential.
The standard shear approximation for quads is generally adequate
for fitting the data, but not for recovering correct parameter
values.  The essential point is that the environment-related errors
are not random errors but systematic biases, such as overestimates
of lens galaxy velocity dispersions, the Hubble constant, and
$\Omega_\Lambda$, and underestimates of the image magnifications.

These systematic effects help resolve one long-standing puzzle
in lensing, but seem to aggravate two others.  The solved problem
is the high number of quad lenses relative to doubles
\citep[also see][]{cohn}.  We have shown that neglecting environment
can cause models to significantly underestimate the quad/double
ratio; accounting for environment can increase the ratio to near the
observed value.

The first unsolved problem involves lensing and the Hubble constant.
Analyses of lens time delays that invoke what we think we know about
galaxy dark matter halos yield $H_0 \sim 50$~km~s$^{-1}$~Mpc$^{-1}$,
which is low compared with the conventional value $H_0 \sim 70$
\citep[see][and references therein]{csk02,csk03}.  We find that
that poor knowledge of environments causes an overestimate of $H_0$,
so accounting for environment would {\it worsen\/} the problem.
The extent of this problem is actually unclear, because some other
lensing analyses yield $H_0 \sim 70$ \citep[e.g.,][]{kt1608,saha04}.
In any case, there is a discrepancy remaining to be resolved, and
the role of environments cannot be ignored.

The second puzzle involves lensing constraints on $\Omega_\Lambda$.
Traditional lensing analyses have tended to produce low values of
$\Omega_\Lambda$, such as the oft-quoted bound $\Omega_\Lambda<0.66$
at 95\% confidence \citep{csk96}, that are in marginal conflict
with the concordance cosmology.  We find that poor knowledge of
environments leads to an overestimate of $\Omega_\Lambda$, which
would seem to worsen the problem.  However, updated analyses have
revised the lensing results upwards, with best-fit values in the
range $\Omega_\Lambda \approx 0.72$--$0.78$ \citep{mitchell}.  In
this case, correcting for environment might even resolve any
discrepancies that may remain between lensing and other methods.

Apart from a general desire to make lensing analyses accurate,
there is new impetus to control the systematics.  \citet{linder}
recently pointed out that if lens models can be made accurate at
the $\sim$1\% level then strong lensing can probe the dark energy
equation of state in a manner complementary to the other probes now
being discussed (such as supernovae, the microwave background, and
weak lensing).  We have argued that as long as lens environments
are poorly understood there is no way for lensing to achieve this
kind of accuracy.  But if environments receive the kind of detailed
study they deserve, the prospects are good for lensing to participate
in the new era of precision cosmology.

\vspace{1cm}
\acknowledgements
We thank Iva Momcheva and Kurtis Williams for valuable discussions,
and for allowing us to use new data for PG~1115+080 in advance of
publication.  We thank Joanne Cohn for interesting discussions.
C.R.K.\ is supported by NASA through Hubble Fellowship grant
HST-HF-01141.01-A from the Space Telescope Science Institute,
which is operated by the Association of Universities for Research
in Astronomy, Inc., under NASA contract NAS5-26555.
A.I.Z.\ is supported by NSF grant AST-0206084 and NASA LTSA grant
NAG5-11108.


\end{document}